\documentclass{iitpressproc}

\usepackage{graphicx}
\usepackage{amsmath}

\usepackage[sort&compress]{natbib}
\bibpunct{[}{]}{,}{n}{}{}

\begin{document}

\title{Minimum-bias angular and trigger-associated correlations from 200 GeV p-p collisions: jets, flows, centrality and the underlying event}

\author{D.~Prindle for the STAR collaboration
\address{CENPA 354290, University of Washington, Seattle, USA}\\[2ex]
}

\maketitle

\begin{abstract}
The mechanisms leading to the hadronic final state of high-energy proton-proton collisions remain an unresolved issue at the RHIC and LHC. A
substantial contribution to the hadronic final state from minimum-bias (MB) jets is dominated by non-perturbative processes and may
provide the common base for any high-energy dijet. Observation of a same-side (on azimuth)``ridge'' in LHC p-p collisions suggests
to some that hydrodynamic flows may play a role in that small  system at higher energies. The issue of p-p centrality vs triggered jets has
emerged in the context of gluon transverse distributions in the proton inferred from DIS data. Attempts have been made to isolate
and study the underlying event (UE) complementary to triggered dijets, and it is suggested that multiple parton interactions may
contribute to the UE. 

Reference [1] considered theoretical and experimental results for UE systematics and p-p centrality in the context of a
two-component (soft+hard) model derived from single-particle $p_t$ spectrum $n_{ch}$ systematics. The study concluded that there may be a
substantial contribution to the UE from the triggered dijet and that p-p centrality is not controlled significantly by a jet
trigger condition (if p-p centrality is relevant at all). Further study of two-particle correlations in p-p collisions was called
for, particularly the $n_{ch}$ dependence of MB correlations.

We report a comprehensive study of MB (no $p_t$ cuts) angular correlations and trigger-associated (TA) $y_t$ correlations (transverse
rapidity $y_t = \ln[(m_t + p_t)/m_\pi])$ from 200 GeV p-p collisions. Angular correlations are characterized by 2D model fits that
accurately distinguish among proton dissociation structure (soft), jet-related structure (hard) and a nonjet azimuth quadrupole.
All angular correlations are simply represented by a (2+1)-component model. The hard and quadrupole component scale simply with the
soft-component multiplicity $n_s$, clarifying the role of centrality and the eikonal model in p-p collisions. 2D TA correlations
project to a marginal 1D trigger spectrum that can be simply predicted from $p_t$ spectrum $n_{ch}$ dependence. 2D TA distributions can
then be processed to reveal MB jet fragment (hard component) systematics comparable to measured fragmentation functions.
Hard-component azimuth dependence relative to the trigger relates to UE studies. From TA analysis we can establish the kinematic
limits of jet fragment production in p-p collisions. 
\end{abstract}

\section{Introduction}

In this paper we present an analysis of particle production in proton-proton collisions at a center of mass energy of 200 GeV.
The data were recorded by the STAR detector at RHIC in 2009 during a low luminosity running period which enabled an
efficient minimum-bias (MB) trigger with very little, if any, pileup.
We have enough statistics to divide the sample into six multiplicity bins ranging from a mean multiplicity of about two
to 27 in two units of rapidity.

In section 2 we recap a few features of the 1D two component spectrum model (TCM) that we use in understanding the
MB two-particle angular correlations presented in section 3.
The angular correlations are described with a few components.
In section 4 we extend the 1D TCM to 2D trigger-associated (TA) correlations.
The TA model describes the trigger spectra and associated soft component very well allowing us to subtract the TA model soft component
and study the data hard component in detail.
The hard component is discussed in section 5, focusing on toward, transverse (trans) and away azimuth regions as conventionally defined in UE studies.
We summarize in section 6.

\section{1D Two-component model}

The two-component spectrum model was developed to understand $y_t$ (or $p_t$) spectra in proton-proton collisions.\cite{bib2}
It was observed that the spectrum shape can be decomposed into soft, $S_0(y_t)$, and hard, $H_0(y_t)$, components where
the soft component is defined by the spectrum shape as the event multiplicity goes to zero.
The component shapes are independent of multiplicity but the ratio of their amplitudes depends on multiplicity: $n_h/n_s \propto n_s \approx n_{ch}$
where $n_s$ and $n_h$ are the soft and hard components of $n_{ch}$ within the acceptance $\Delta\eta$.
In analogy with A-A collisions we can think of the soft component as being related to participants (low-x gluons for p-p) and the hard component to
binary parton collisions.
In a Glauber calculation of A-A collisions the numbers of participants and binary collisions depend on the impact parameter and
we find $N_{bin} \propto N_{part}^{4/3}$.
For proton-proton collisions we find $n_h \propto n_{s}^2$; a Glauber description doesn't work.

\section{Angular correlations of particles produced in p-p collisions}

\begin{figure}[h]
\centering
\includegraphics[width=0.95\columnwidth]{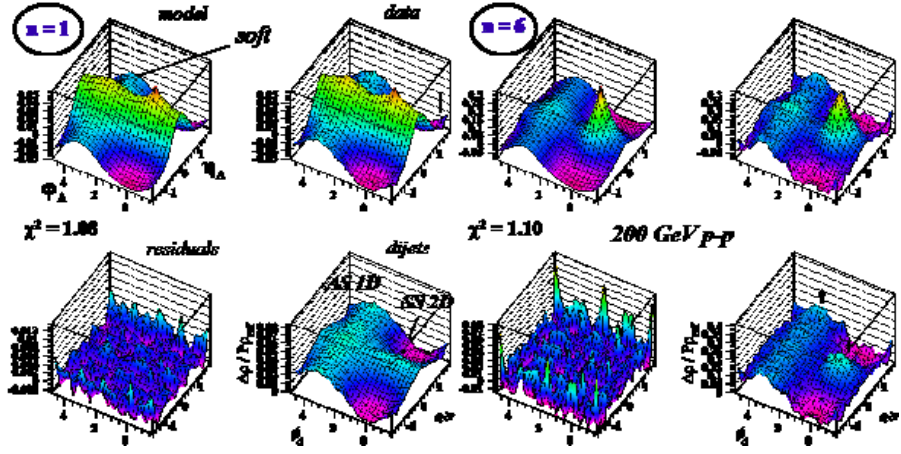}
\caption{MB angular correlations.
The four panels on the left are for low multiplicity and the four panels on the right for high multiplicity events.
In each group the upper left quadrant shows the model, the upper right quadrant shows the data and the lower left shows the residuals.
In the lower right quadrant the 1D Gaussian and HBT/$e^+e^-$ model components have been removed leaving the dijet and non-jet quadrupole components.\label{AngularFig}}
\end{figure}

We developed a MB method to analyze two particle angular correlations to study A-A collisions without imposing preconceived ideas.\cite{porter2}\cite{porter3}\cite{axialci}\cite{anomalous}
In this method we take all pairs of particles and project out $\phi_1 + \phi_2$ and $\eta_1 + \eta_2$ which loses no information if we have
rotational symmetry in $\phi$ (which is true) and translational symmetry in $\eta$ (which is a reasonable approximation at mid-rapidity).
Integrating over $p_t$ we end up with a two dimensional correlation on $\phi_\Delta \equiv \phi_1 - \phi_2$ and
$\eta_\Delta \equiv \eta_1 - \eta_2$.
This analysis works for even very low multiplicity events so we can apply it to proton-proton collisions.
In Fig.~\ref{AngularFig} we show the angular correlations for the lowest and highest multiplicity bins.
In each case the upper left quadrant is the model, the upper right quadrant is the data and the lower left quadrant is the residuals.
The model consists of a 2D Gaussian at the origin, a narrow 2D exponential at the origin, an away-side 1D azimuth dipole and a 1D Gaussian on $\eta_\Delta$.
There is a significant improvement in the fit when we include a nonjet quadrupole $\cos(2\phi_\Delta)$ component which is clearly required in A-A correlations.
By examining the $p_t$ and charge dependence of these components we find that the narrow 2D exponential is due to HBT and $\gamma$ conversion to $e^+e^-$ pairs,
the 1D Gaussian on $\eta_\Delta$ is consistent with soft particle emission from the beam remnants and the same-side 2D Gaussian is due to intra-jet
correlations while the away-side dipole is due to inter-jet correlations.
The lower right quadrants show the data with the 1D Gaussian on $\eta_\Delta$ and the  HBT/$e^+e^-$ components removed
leaving the jet structure and non-jet quadrupole.

\begin{figure}[h]
\centering
\includegraphics[width=0.45\columnwidth]{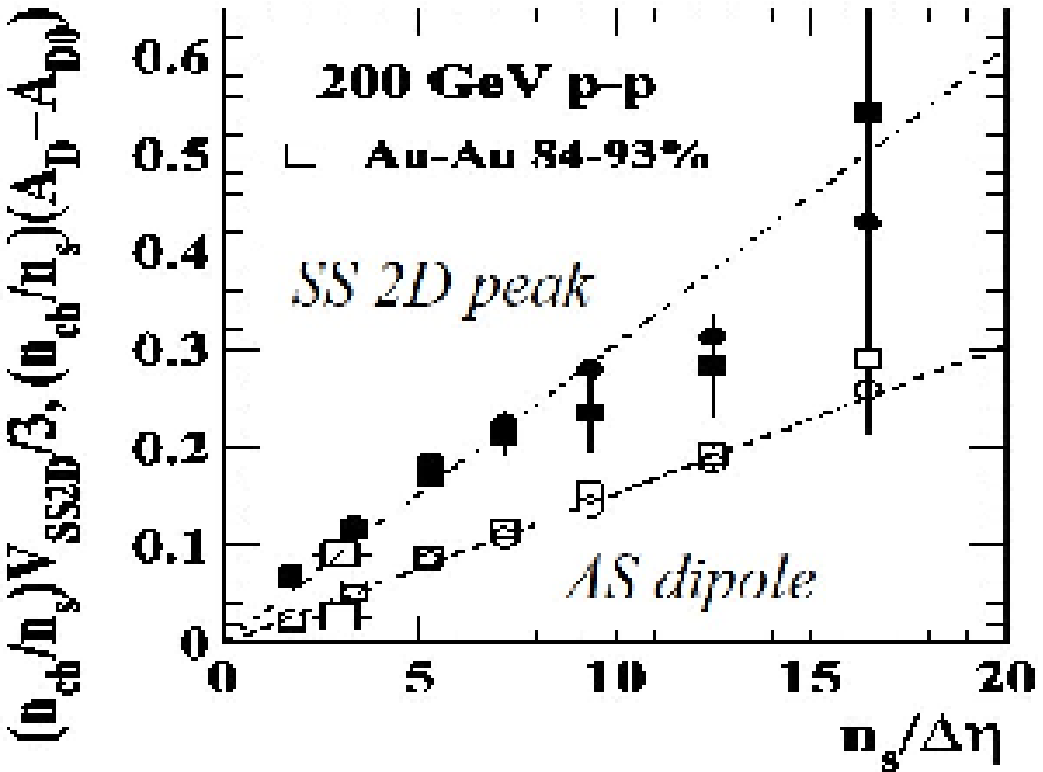}%
\includegraphics[width=0.45\columnwidth]{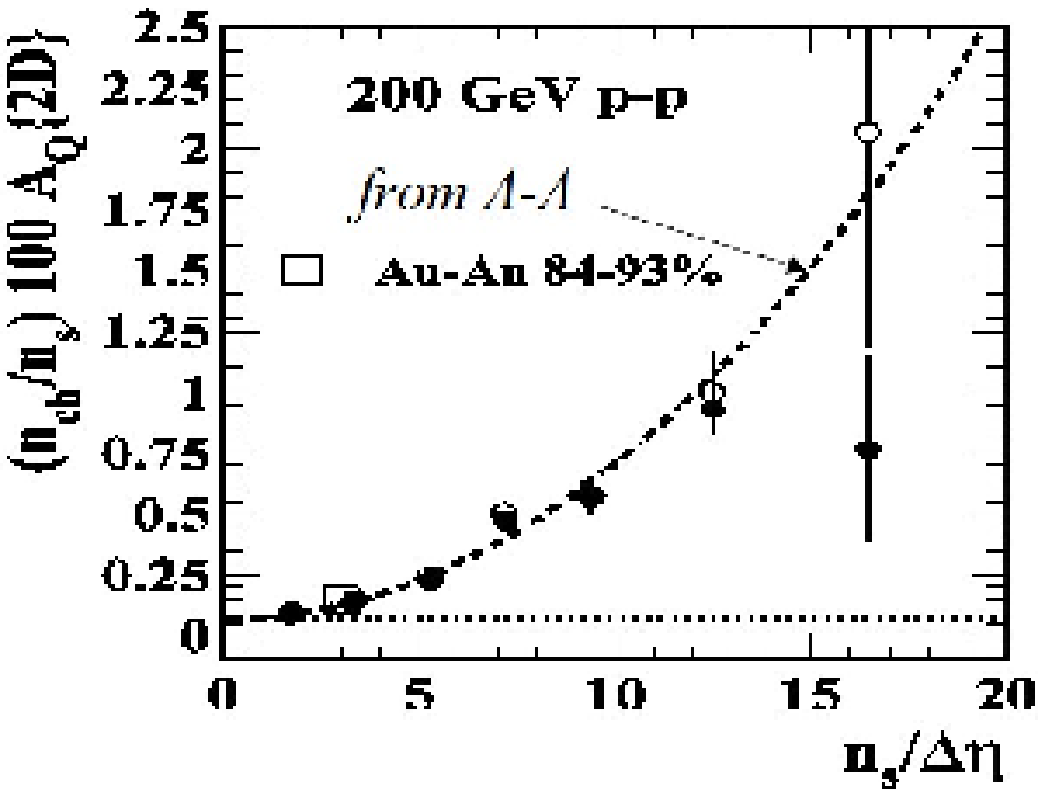}
\caption{The left panel shows the volume of the same-side 2D Gaussian ($V_{SS2D}$) and the away-side dipole amplitude,
both scaled by $n_{ch}/n_s$.
The away-side dipole amplitude has been adjusted by subtracting $A_{D0}$ to account for global transverse momentum conservation.
$n_{ch}$ and $n_s$ are the total multiplicity and soft component of the multiplicity within the acceptance, $\Delta\eta$.
The right panel is the nonjet quadrupole amplitude, $A_Q$, also scaled by $n_{ch}/n_s$.
We observe that $n_{ch}$ times each dijet component amplitude is proportional to $n_s^2$
while $n_{ch} A_Q$ is proportional to $n_s^3$.\label{AngularParams}}
\end{figure}

In Fig.~\ref{AngularParams} we show the multiplicity dependence of the dijet amplitudes for the same-side 2D peak and the away-side dipole.
The amplitudes of the dijet structures scale with the square of the multiplicity and are consistent with a QCD dijet total cross section
$\sigma_{dijet} = 2.5$~mb\cite{fragevo}.
The non-jet quadrupole component is interesting.
In A-A collisions we find that $n_{ch} A_Q \propto N_{part} N_{bin} \varepsilon_{opt}^2$.
We assume that in proton-proton collisions $\left < \varepsilon_{opt}^2 \right >$ is non-zero and note that $N_{bin} \propto N_{part}^2$.
If the non-jet quadrupole in proton-proton collisions arises from the same mechanism as in A-A collisions we predict that
\begin{equation}
\left ( n_{ch} / n_s \right ) A_Q \propto n_s^2
\end{equation}
which indeed appears to be the case.
This suggests that we should seriously consider if the non-jet quadrupole in proton-proton collisions is due to the same
physics as in A-A collisions and is unrelated to hydrodynamic flows.

\section{Trigger-associated analysis\label{section4}}

We extend the 1D TCM to a 2D trigger-associated (TA) model to isolate the hard component in $(y_{tt},y_{ta})$
as well as to connect with underlying event (UE) studies.
Here $y_{tt}$ is $y_t$ of the trigger and $y_{ta}$ is $y_t$ of the associated particle.
For a MB analysis we take the track with the highest $y_t$ in the event as the trigger, all other tracks are associated.
Thus we accept all pairs from all jets in the analysis.
A useful constraint on the 2D TA model is that the projection onto the associated particle axis is the 1D single-particle spectrum minus the
trigger spectrum.

The trigger particle may be from an event with no hard component (no dijet in the acceptance) in which case the spectrum is derived from the soft component only.
If the event has a hard component (at least one jet) then the trigger can be due to the soft or hard component depending on which produced
the highest $y_t$ particle of the event.
The void probability $G$ is in either case the probability that no particle appears above $y_{tt}$.
\begin{align}
\rho_{trig}(y_{tt},n_{ch}) = &P_s(n_{ch})G_s(y_{tt},n_{ch})S_o(y_{tt}) + \\
                             &P_h(n_{ch})G_h(y_{tt},n_{ch})F_h(y_{tt})   \notag
\end{align}
where
\begin{equation}
F_h(y_{tt},n_{ch}) = p'_s(n_{ch})S_0(y_{tt}) + p'_h(n_{ch})H_0(y_{tt})
\end{equation}
$P_s$ and $P_h \equiv (1-P_s)$ are probabilities for soft and hard events, $p'_s$ and $p'_h$ are the soft and hard probabilities
given that there is a hard component, $G_s$ and $G_h$ are the
void probabilities and $S_0$ and $H_0$ are the soft and hard components of the 1D TCM.
All functions and probabilities are taken from the 1D TCM and the predicted trigger spectra are in excellent agreement with
the measured trigger spectra.

The 2D TA distribution
\begin{equation}
F(y_{ta},y_{tt},n_{ch}) = T(y_{tt}) A(y_{ta}|y_{tt})
\end{equation}
is the joint probability of a trigger, $T(y_{tt})$, and an associated particle, $A(y_{ta}|y_{tt})$,
where $A$ is the conditional probability of an associated particle being emitted at $y_{ta}$ in an event with trigger $y_{tt}$.
We combine soft and hard components for $F$ according to the 1D TCM.
Complete details are given in reference\cite{TAMath}.

\begin{figure}[h]
\centering
\includegraphics[width=0.95\columnwidth]{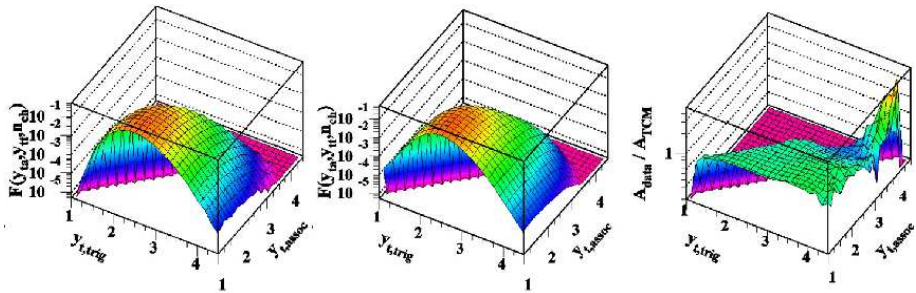}
\caption{The left panel contains TA correlations, the middle panel is the TA model based on the 1D spectrum TCM.
The right panel is the data to TA model ratio of the associated spectrum (A) showing that the soft component is described very well (ratio $\approx 1$) by the TA model for $y_{ta} < 2.5$.
These are for multiplicity bin 3, the other multiplicity bins are equally well described.\label{2DTCM}}
\end{figure}

In the previous paragraphs we have sketched how we extend a 1D TCM spectrum model of proton-proton collisions to a 2D TA model.
In Fig.~\ref{2DTCM} we compare the measured correlations (left panel) with the model (middle panel).
These are for multiplicity bin 3 but other multiplicity bins show similar agreement.
The $y_{t,trig}$ axis of these panels reflect the trigger spectrum.
We are interested in the associated spectrum which we get by dividing the 2D TA correlations by the 1D trigger spectrum, $A\equiv F/T$.
The third panel of Fig.~\ref{2DTCM} is the ratio of data and TA model for A.
The soft component is described very well (ratio $\approx 1$) for $y_{ta} < 2.5$.
We subtract the TCM model soft component from the data to reveal the data hard component in detail.

\section{Hard component of A}

\begin{figure}[h]
\centering
\includegraphics[width=0.95\columnwidth]{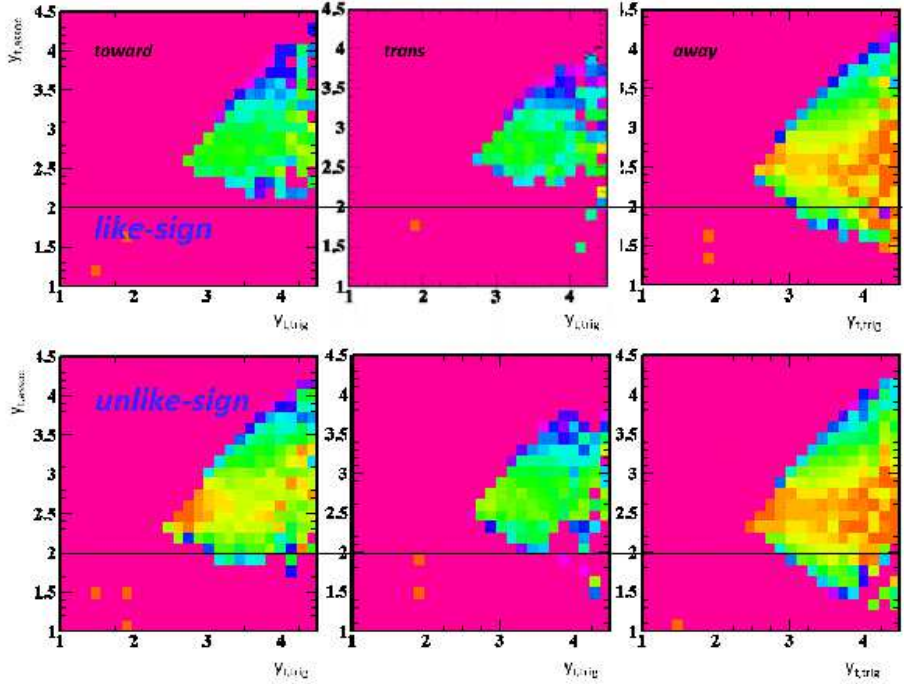}
\caption{Upper row are like-sign pairs of the hard component of the TA correlations. Lower row are unlike-sign pairs.
Left column is toward ($\Delta\phi\leq\pi/3$), middle is trans ($\pi/3<\Delta\phi\leq 2\pi/3$) and right is away ($2\pi/3<\Delta\phi$).
Toward shows charge ordering while the away is charge independent.
There is a significant jet-related yield in trans.
\label{CDHard}}
\end{figure}

In section~\ref{section4} we discussed an extension of the 1D TCM spectrum model to 2D in order to describe MB TA correlations
and we showed this accurately described the trigger spectrum and the soft part of the associated spectrum.
This enables us to isolate the associated hard component of data by subtracting the model soft component.
In Fig.~\ref{CDHard} we show the hard component for like-sign and unlike-sign pairs in the toward ($\Delta\phi\leq\pi/3$),
trans ($\pi/3<\Delta\phi\leq 2\pi/3$) and away ($2\pi/3<\Delta\phi$) regions.
In UE studies the toward and away regions are usually assumed to be dominated by jets and
the trans to be dominated by the UE.
While the trans may have fewer jet-related particles than toward and away there are still a significant number that can't be ignored.
The toward region shows a charge-ordering effect with more unlike-sign pairs than like-sign pairs.\cite{porter3}
The away-side is charge independent and the associated particles extend to lower $y_t$ than the same-side.
This can be understood as due to the initial $k_t$, for back to back particles of equal $y_t$ one will be boosted
to higher energy and become the trigger while the other will be reduced in energy.

\section{Summary}

We have analyzed a very clean sample of 200 GeV proton-proton collisions.
To understand the data we reviewed the characteristics of the soft and hard components of the TCM developed
to understand 1D $y_t$ spectra.
Then we extracted jet-like components and a non-jet quadrupole from the 2D angular correlations on $(\eta_\Delta,\phi_\Delta)$.
We showed the non-jet quadrupole may have the same origin in proton-proton collisions as it does in A-A collisions.
We did a MB TA analysis where the highest $y_t$ in the event is taken to be the trigger.
To understand this we extended the TCM to a 2D TA model which describes the trigger spectrum and soft component of the
data very well.
Subtracting the TA model soft component we see that the remaining hard component exhibits charge ordering for
the toward region but not the away region, as expected for jets.
The trans region exhibits a significant jet structure.

\section*{Acknowledgments}
  This work is supported in part by the Office of Science in the U.S. DOE under grant DE-FG03-97ER41020.


\bibliographystyle{apsrev}
\bibliography{prindle-ismd13}    

\end{document}